# Variations of flaring kernel sizes in various parts of the Hα line profile

K. Radziszewski and P. Rudawy

**Abstract** We analyze the temporal variations of the sizes and emission intensities of thirty-one flaring kernels in various parts of the Hα line profile. We have found that the areas of all kernels decrease systematically when observed in consecutive wavelengths toward the wings of the Hα line, but their areas and emission intensity vary in time. Our results are in agreement with the commonly accepted model of the glass-shaped lower parts of the magnetic flaring loops channelling high energy variable particle beams toward the chromospheric plasma.

High time resolution spectral-imaging data used in our work were collected using the *Large Coronagraph* and *Horizontal Telescope* equipped with the *Multi-channel Subtractive Double Pass Spectrograph* and the *Solar Eclipse Coronal Imaging System* (MSDP-SECIS) at the Białków Observatory of the University of Wrocław, Poland.

1. Introduction

Satellite-based observations of solar flares collected in the hard X-ray domain usually reveal some hard X-ray sources located near the feet of flaring loops (they form the so-called "foot-point HXR sources"), while co-temporal ground-based observations recorded in visible wavelengths (for example in strong chromospheric hydrogen Hα line, 656.3 nm) reveal bright compact and/or extended emission sources located in the closest vicinity of the X-ray sources. It is commonly accepted that both types of sources are located in the feet of the flaring magnetic loops, where a relatively dense and cold chromospheric plasma is heated mainly by charged, high-energy particles streaming down along the loops from primary energy sources located somewhere close to the tops of the loops. The crucial parameters describing the properties of the energetic particle beams, like total energy flux or energy distribution of the particles are usually highly variable in time.

K. Radziszewski , P. Rudawy
Astronomical Institute of the University of Wrocław, 51-622 Wrocław, ul. Kopernika 11, Poland
e-mail: radziszewski@astro.uni.wroc.pl

Due to the very fast change of the magnetic β parameter (*i.e.* the ratio of the plasma pressure ($p=nk_BT$) to the magnetic pressure ($p_{mag}=B^2/2\mu_0$), where: $n$ is the number density, $k_B$ is the Boltzmann constant, $T$ is the temperature, $B$ is the strength of the magnetic field and $\mu_0$ is the magnetic permeability, see Cravens, 1997 for details) between the lower corona, chromo-sphere and photosphere, lower parts of the flaring magnetic loops should shrink significantly, and thus they are thought to be roughly glass-shaped (*e.g.* Gabriel, 1976; Foukal, 1990; Stix, 1989; Aschwanden, 2006). While various parts of the Hα line are formed at different depths in the chromosphere (Vernazza, Avrett and Loeser, 1973, 1981; Kašparová and Heinzel, 2002; Berlicki and Heinzel, 2004), taking into account all obvious factors like a hetero-geneous vertical stratification of the plasma, strong bulk and turbulent plasma motions inside the flaring loops and many more, an emission recorded in a particular part of the Hα line profile enables a crude determination of the precipitation depths of the non-thermal electrons, as well as an evaluation of sizes of the emitting flaring kernels at various levels (it means vertical variations of the cross-sections of the flaring loops).

In our previous work (Radziszewski and Rudawy, 2008) we presented results of spectrophotometric investigations of instantaneous sizes of seven Hα flaring kernels, when measured in various parts of the Hα line profile at selected moments. We have observed that the areas of the investigated individual kernels decreased systematically when observed simultaneously in consecutive wavelengths toward the wings of the Hα line, in agreement with the theoretical models of magnetic fields and chromospheric emission mentioned before. While the wings of the Hα line arises from dipper layers of the chromosphere than the line core, due to the cone-shaped lower part of the loop the actual size of the flaring kernel will depend on the wavelength in the frame of the Hα line profile.

In the present paper we describe extended investigations of the temporal variations of the sizes of the 31 individual flaring kernels, observed between 2003 and 2005, using for each kernel up to 10 thousand measurements with high-time resolution (up to 40 ms).

The observational data are described in Section 2, the data reduction is described in Section 3, while the results are presented in Section 4 and the discussion of the results and conclusions are presented in Section 5.

## 2. Observations

High-time resolution, long sets of so-called *spectra-images* (*i.e.* two dimensional images of the observed solar region convolved with the Hα line spectra, see Mein, 1977 for details) of the flaring kernels were collected with the *Large Coronagraph* (LC) and the *Horizontal Telescope* (HT) equipped with *the Multi-channel Double Pass* (MSDP) imaging spectrograph and the *Solar Eclipse Coronal Imaging System* (SECIS) at the Białków Observatory of the University of Wrocław, Poland.

The LC has a 53 cm diameter main objective; its effective focal length is equal to 1450 cm. The spatial resolution of the instrument, usually limited by visual conditions, is about 1 arcsec. The HT has a compact Jensch-type coelostat with 30 cm mirrors; its main objective has an aperture of 15 cm and a 5 m focal length. The MSDP spectrograph has a rectangular (2D) entrance window, which covers an equivalent area of 942 × 119 arcsec$^2$ on the Sun when lighted by the HT or an equivalent area of 325 × 41 arcsec$^2$ on the Sun when lighted by the LC and a nine channel prism-box (Mein, 1991; Rompolt *et al.*, 1994). The spectra-images created by the MSDP spectrograph were recorded with the fast CCD camera of SECIS system (512 × 512 px$^2$, 1 px$^2$=1 arcsec$^2$, up to 70 images per second) (see Phillips *et al.*, 2000; Rudawy *et al.*, 2004 for details). After standard numerical reduction of each spectra-image we obtained Hα spectra for all pixels inside the field of view (in a range of ±0.12 nm from the line centre) and quasi-monochromatic images of the whole field of view, reconstructed in freely-chosen wavelengths in the Hα line profile bandwidth range (in the present work separated in wavelengths by 0.02 nm from each other). Extended information concerning the MSDP-SECIS system and data reduction details are given in papers by Radziszewski *et al.* (2006) and by Radziszewski, Rudawy, Phillips (2007).

**Table 1** The list of the analyzed high-cadence spectral observations of the Hα flaring kernels.

| Date | Obs. period [UT] | AR | Location | GOES class | Cadence [ms] | Tele-scope | Hα kernels |
|---|---|---|---|---|---|---|---|
| 2003 Jul 16 | 15:57:45-16:06:05 | 10410 | S10 E28 | C1.2 | 50 | LC | K1, K2 |
| 2004 May 03 | 07:24:15-07:34:18 | 10601 | S08 W54 | B2.5 | 60 | LC | K9, K10 |
| 2004 May 05 | 08:38:30-08:46:50 | 10605 | S10 W10 | B3.1 | 50 | HT | K11 |
| 2004 May 05 | 11:44:35-11:51:15 | 10605 | S10 W10 | B5.4 | 40 | HT | K12 |
| 2004 May 21 | 05:44:08-05:50:48 | 10618 | S10 E55 | C2.0 | 40 | LC | K13 |
| 2004 May 21 | 10:25:26-10:30:06 | 10618 | S10 E55 | B7.0 | 40 | HT | K14 |
| 2005 Jan 17 | 08:00:59-08:11:59 | 10720 | N13 W29 | X3.8 | 66 | LC | K15, K16, K17, K18 |
| 2005 Jul 12 | 07:53:10-08:10:20 | 10786 | N09 W68 | C8.3 | 50 | HT | K19, K20 (2 series) |
| 2005 Jul 12 | 10:00:44-10:09:04 | 10786 | N09 W68 | C2.3 | 50 | HT | K21 |
| 2005 Jul 12 | 12:10:15-12:18:34 | 10786 | N09 W68 | C1.5 | 50 | HT | K22, K23 |
| 2005 Jul 12 | 13:02:11-13:10:30 | 10786 | N09 W68 | M1.0 | 50 | HT | K25, K26 |
| 2005 Jul 12 | 15:27:57-15:39:01 | 10786 | N09 W68 | C2.3 | 66 | HT | K30, K31 |
| 2005 Jul 12 | 15:27:57-15:39:01 | 10786 | N09 W68 | C2.3 | 66 | HT | K27, K28, K29 |
| 2005 Jul 12 | 15:39:34-15:50:39 | 10786 | N09 W68 | M1.5 | 66 | HT | K33 |
| 2005 Jul 13 | 08:15:04-08:26:09 | 10786 | N11 W79 | C2.7 | 66 | HT | K35, K36, K37 |
| 2005 Jul 13 | 10:05:40-10:16:45 | 10786 | N11 W79 | C1.6 | 66 | HT | K38 |
| 2005 Jul 13 | 12:04:40-12:15:45 | 10786 | N11 W79 | M3.2 | 66 | HT | K40 |
| 2005 Aug 26 | 11:42:21-11:50:41 | 10803 | N12 E53 | C2.1 | 50 | LC | K42 |

Between July 2003 and August 2005 we observed many solar flares having well visible individual flaring kernels. For each kernel we recorded from 10 up to 20 thousand spectra-images with time resolution from 75 ms (slightly more than 13 images per second) up to 40 ms (25 images per second), depending on the intensity of the light beam.

We selected 31 flaring kernels observed during 18 solar flares in seven active regions. All kernels are labelled in accordance with the nomenclature of the kernels applied in our previous paper by Radziszewski, Rudawy and Phillips (2011), which includes also a full list of the observed flares. For two kernels, K19 and K20, we collected data in two consecutive observing periods. For convenience, we analyze both data sets separately. Thus the total number of the described events is equal to 33. The detailed list of the investigated kernels is given in Table 1.

In Figure 1 we present, as a representative example, the 2D quasi-monochromatic image of the flare (kernels K19 and K20) recorded in the Hα line centre in NOAA 10786 active region on July 12, 2005.

## 3. Data reduction

In principle the MSDP imaging spectrograph equipped with the SECIS fast recording system is very suitable for studies of the fast variations of the line profiles emitted by individual Hα flaring kernels and their sizes. However, ground-based observations of the Sun are strongly affected by the highly variable atmospheric seeing, causing strong and variable in time deformations of the apparent shapes of the recorded structures. Thus, we applied special procedures in order to correct shifts of the telescope pointing, variations of the atmospheric transmittance and seeing-induced distortions of the images (see Radziszewski, Rudawy and Philips (2007) for details).

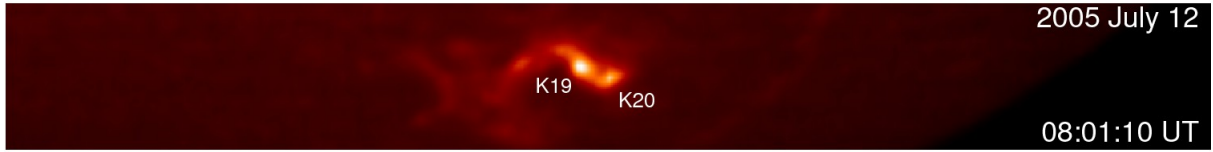

**Figure 1** Hα line centre image of the solar flare on 2005 July 12 observed with the HT-MSDP-SECIS system at Białków Observatory. The Hα emission kernels are marked K19 and K20 (see main text for details). The field of view is equal to 942×119 arcsec$^2$.

Using proprietary automatic codes we established for all investigated kernels their temporal variations of area (or its equivalent radius defined as the radius of the circle having the same area) and brightness versus wavelength. Both parameters were calculated for each kernel and for each spectra-image simultaneously in 11 wavelengths, up to ±0.1 nm from the Hα line centre. We took into account all pixels encompassed by isophotes selected arbitrary at a level of 75% of the net maximum brightness of the investigated kernel (i.e. with subtracted brightness of the relevant quiet chromosphere). The areas of the emitting structures are evaluated in square pixels only in order to avoid any uncertainties caused by projection effects (in the centre of the solar disk 1 px$^2$ = 3.1 arcsec$^2$ (5×10$^6$ km$^2$) when observed with the HT and 1 px$^2$ = 1.1 arcsec$^2$ (~6×10$^5$ km$^2$) when observed with the LC, respectively).

## 4. Results

The detailed analysis of the collected data shows that the flaring kernel areas observed in the Hα line centre are significantly greater than the areas observed in the Hα line wings and the kernel areas decreased systematically when observed from the centre of the Hα line towards its wings. Most of the observed kernels did not show any asymmetries of the area vs. wavelength curves, *i.e.* the maxima of these curves appeared for the central part of the Hα line profile. However, for some kernels we recorded asymmetries of these curves, *i.e.* the maxima of the curves were shifted toward the shorter or longer wavelengths, usually by 0.02-0.04 nm from the line centre.

The observations of 15 kernels (45% of the whole set of data; thereafter called Group I) started well before any noticeable increase of the soft X-ray flux recorded by GOES, when the state of the Hα emitting plasma was not yet changed due to the energy delivered by non-thermal electrons (plasma evaporation), or the magnitude of the changes was rather limited. Even before the impulsive phase some non-thermal electrons could be present in the flaring loops, causing some faint emission (Siarkowski, Falewicz and Rudawy, 2009; Falewicz, Siarkowski and Rudawy, 2011). In opposite, for 18 kernels (55%) the observations started during the impulsive or even maximum phases of the relevant solar flare (thereafter called Group II). In most cases increases of the observed Hα emissions were well correlated in time with relevant increases of hard X-ray emissions recorded by the *Reuven Ramaty High-Energy Solar Spectroscopic Imager* (RHESSI) satellite (Radziszewski, Rudawy and Phillips, 2011).

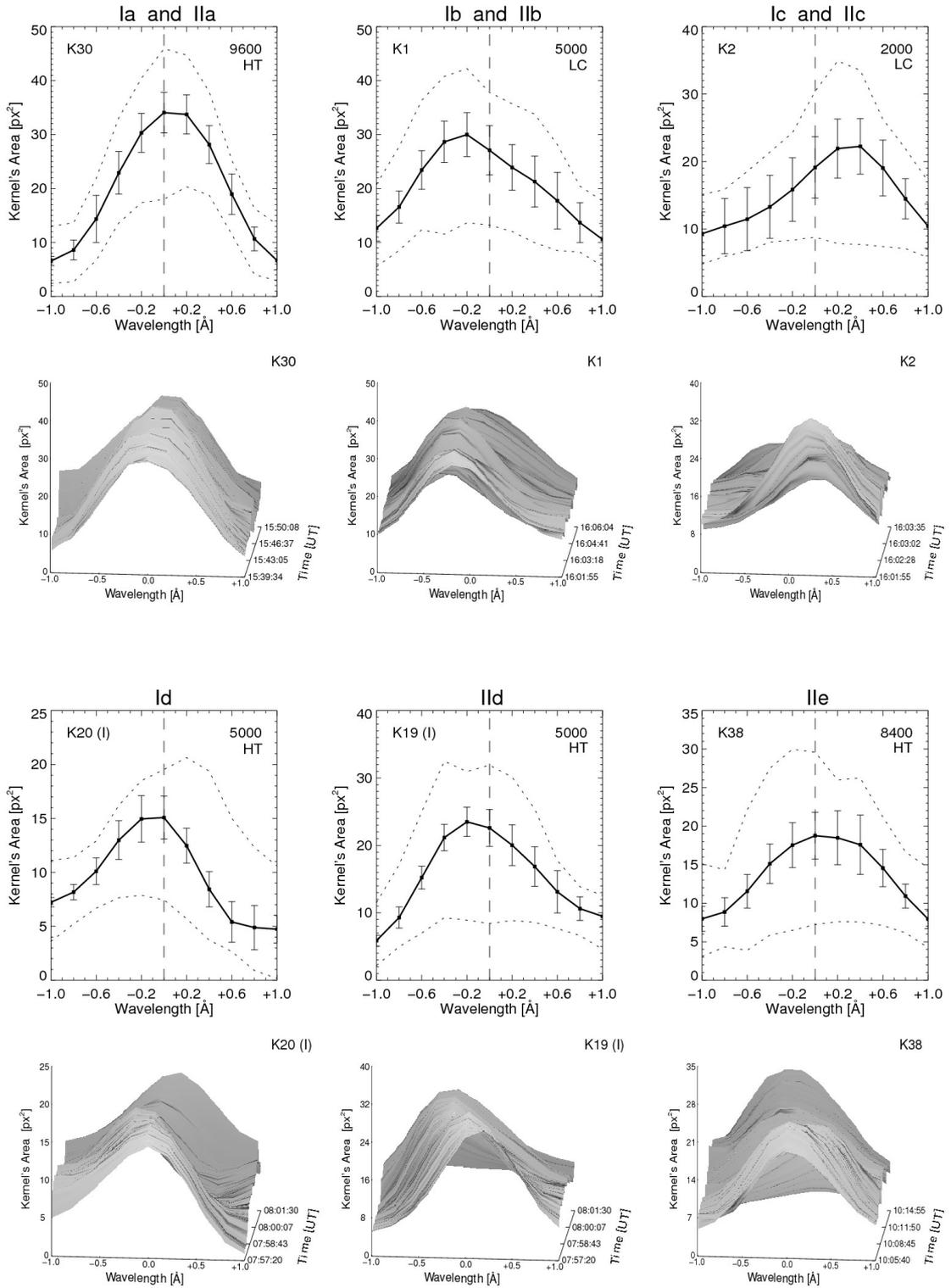

**Figure 2** Six representative examples showing the time variation of the areas vs. wavelength of the analyzed Hα flaring kernels. For each kernel the upper plots show mean (solid line), maximum and minimum (dotted lines) instant areas measured in 11 wavelengths, up to ±0.1 nm from the Hα line centre. The groups to which the kernels belong are indicated in the titles. The numbers of the analyzed (and averaged) measurements for each wavelength separately are given in the upper-right corners of the relevant figures. HT means data collected with the *Horizontal Telescope*, LC with the *Large Coronagraph*. Lower plots show time variations of the areas vs. wavelength. The wavelengths are counted from the Hα line centre. The areas of the kernels are given in square pixels (in the centre of the solar disk 1 px$^2$ = 3.1 arcsec$^2$ (5×10$^6$ km$^2$) when observed with the HT and 1 px$^2$ = 1.1 arcsec$^2$ (~6×10$^5$ km$^2$) when observed with the LC, respectively). Error bars show standard deviations of the measurements. Groups Ia and Ib are represented by K30, groups Ib and IIb by K1; groups Ic and IIc by K2; group Id by K20; group IId by K19 and group IIe by K38, respectively.

**Table 2** Group and number of the flaring kernels.

| Group | Type | Number | Description |
|---|---|---|---|
| I (observations started before the impulsive phase) | Ia | 9 | maximum areas observed constantly in the Hα line centre |
| | Ib | 2 | maximum areas observed constantly in the blue wing of the Hα line |
| | Ic | 2 | maximum areas observed constantly in the red wing of the Hα line |
| | Id | 2 | maximum areas observed in the blue wing of the Hα line at the beginning drifting later to the red wing |
| II (observations started during the impulsive phase) | IIa | 10 | maximum areas observed constantly in the Hα line centre |
| | IIb | 1 | maximum areas observed constantly in the blue wing of the Hα line |
| | IIc | 1 | maximum areas observed constantly in the red wing of the Hα line |
| | IId | 3 | maximum areas observed in the red wing of the Hα line at the beginning drifting later to the blue wing |
| | IIe | 3 | maximum areas observed in the Hα line centre at the beginning drifting later to the blue or to the red wing of the Hα line |

From the 15 kernels of Group I, nine kernels (60% of Group I) had maximum areas observed in the Hα line centre during the whole recorded period of their evolution (labelled as type Ia), 2 kernels had maximum areas observed constantly in the blue wing of the Hα line (labelled Ib), 2 kernels had maximum areas observed constantly in the red wing of the Hα line (labelled Ic). For 2 kernels the maximum areas observed in the blue wing of the Hα line at the beginning drifted later to the red wing (labelled Id).

From the 18 kernels of Group II, 10 kernels (56% of Group II) had maximum areas observed in the Hα line centre during the whole recorded period of their evolution (labelled IIa), 1 kernel had maximum area observed constantly in the blue wing of the Hα line (labelled IIb) and also 1 kernel had maximum area observed constantly in the red wing of the Hα line (labelled IIc). For 3 kernels the maximum areas observed in the red wing of the Hα line at the beginning drifted later to the blue wing (labelled IId) while 3 other kernels had maximum areas observed in the Hα line centre at the beginning drifted later to the blue or to the red wing of the Hα line (labelled IIe). These results are summarized in Table 2. Representative examples showing the time variation of the areas *vs.* wavelength of the different groups of the Hα flaring kernels are shown in Figure 2, while the plots for all kernels are available as online material.

The displacements of the area vs. wavelength could be explained as the result of macroscopic motions of the emitting plasma (up or down) along the magnetic loops. For example, the flaring kernels K1 and K2, observed during the C1.2 GOES class solar flare on July 16, 2003, were located in opposite feet of the loop (see Radziszewski and Rudawy, 2008) and they had the greatest emitting areas visible in the blue and red wing of the Hα line, respectively. Due to the projection effects, both shifts are consistent with emissions of the ascending plasma.

We also noticed that the sizes of the kernels measured in a particular wavelength varied with time. The variations recorded in various wavelengths were roughly similar but not strictly correlated. The time-scale of the variations was usually of the order of a few dozens of seconds. Six representative examples of the time variations of the area vs. wavelength are presented in Figure 2 (kernel K30 presents types Ia and Ib, K1 types Ib and IIb, K2 types Ic and IIc, K20 type Id, K19 type IId and K38 type IIe). For each representative kernel we present mean, maximum and minimum instant areas measured in 11 wavelengths (up to ±0.1 nm from the Hα line centre) and time variations of the area vs. wavelength. The lengths of the analysed time series are given in the figures (in the upper-right corners).

For all the investigated flaring kernels the maxima of the area vs. wavelength occurred in the range of ±0.04 nm from the Hα line centre. For most of the kernels visible areas measured in both wings of the Hα line (at about ±0.1 nm from the line centre) were less than 50% of the maximum area. The uncertainties of the mean areas measured for each particular wavelength were estimated as the standard deviations of their temporal variation after removing long-period trends (smoothed with a 60 points (*i.e.* 3 seconds for 50 ms cadence) box-car filter; the lengths of the analyzed time series vary between 1200 up to 10000 images). The errors are much smaller than the variations of the areas vs. wavelengths (see Figure 2 and Figures 3, 4, and 5 in the online materials).

5. Discussion and conclusions

Taking advantage of the unique abilities of the MSDP imaging spectrograph we directly recorded the high-cadence temporal variations of sizes of the 31 Hα flaring kernels during the impulsive phases of their associated flares, measured simultaneously in numerous wavelengths in a range of ±0.1 nm from the Hα line centre with very high time resolution (40-66 ms).

The detailed analysis of the collected spectra-images of the flaring kernels shows that the areas of the kernels decreased significantly and systematically from the Hα line centre toward their wings. However, for some flaring kernels we recorded asymmetries of the area vs. wavelength, most noticeable during the impulsive phase of the flare, i.e. their greatest emitting areas were detected in the blue or red wings of the Hα line. However, the observed shifts could be explained as Doppler-shifts caused by ascending and emitting plasma. Taking into account all limitations imposed by the undoubtedly unstable and turbulent plasma distribution inside the flaring loops, the sizes of the flaring kernels seem to decrease toward deeper layers, in a good qualitative agreement with an expected glass-like shape of the lower part of the flaring loops (see also Radziszewski and Rudawy, 2008). We also noticed that the sizes of the kernels measured in particular wavelengths varied with time. The variations recorded in various wavelengths were not identical and not correlated in time. The time-scale of the variations was of the order of a few dozens of seconds. These changes could be caused by obviously very turbulent bulk motions of the plasma inside the flaring kernel during the impulsive phase, caused by variable heating beams of non-thermal particles. We have not found any significant correlation between the GOES class of the flares and the temporal evolution of the areas of the flaring kernels.

**References**

Aschwanden, M. J.: 2006, Physics of the Solar Corona, Springer Verlag, 127
Berlicki, A., Heinzel, P.: 2004, Astron. Astrophys. 420, 319
Cravens, T. E.: 1997, Physics of Solar System Plasmas, Cambridge, 110
Falewicz, R., Siarkowski, M., Rudawy, P.: 2011, Astrophys. J. 733, 37
Foukal, P.: 1990, Solar Astrophysics, John Wiley and Sons Inc., 364
Gabriel, A. H.: 1976, Roy. Soc. Phil. Trans. Ser. A. 281, 339

Kašparová, J., Heinzel, P.: 2002, Astron. Astrophys. 382, 688
Mein, P.: 1977, Solar Phys., 54, 45
Mein, P.: 1991, Astron. Astrophys. 248, 669
Phillips, K. J. H., Read, P. D., Gallagher, P. T., Keenan, F. P., Rudawy, P., Rompolt, B., Berlicki, A., Buczylko, A., Diego, F., Barnsley, R., et al.: 2000, Solar Phys. 193, 259
Radziszewski, K., Rudawy, P., Phillips, K. J. H., Dennis, B. R.: 2006, Adv. Space Res. 37, 1317
Radziszewski, K., Rudawy, P., Phillips, K. J. H.: 2007, Astron. Astrophys. 461, 303
Radziszewski, K., Rudawy, P.: 2008, Ann. Geophys. 26, 2991
Radziszewski, K., Rudawy, P., Phillips, K. J. H.: 2011, Astron. Astrophys. 535, A123
Rompolt, B., Mein, P., Mein, N., Rudawy, P., Berlicki, A.:1994, JOSO Annual Report 1993, 87
Rudawy, P., Phillips, K. J. H., Gallagher, P. T., Williams, D., Rompolt, B., and Keenan, F. P.: 2004, Astron. Astrophys., 416, 1179
Siarkowski, M., Falewicz, R., Rudawy, P.: 2009, Astrophys. J. 705, L143
Stix, M.: 1989, The Sun, An Introduction, Springer Verlag, 274
Vernazza, J. E., Avrett, E. H., Loeser, R.: 1973, Astrophys. J. 184, 605
Vernazza, J. E., Avrett, E. H., Loeser, R.: 1981, Astrophys. J. Suppl. Ser. 45, 635

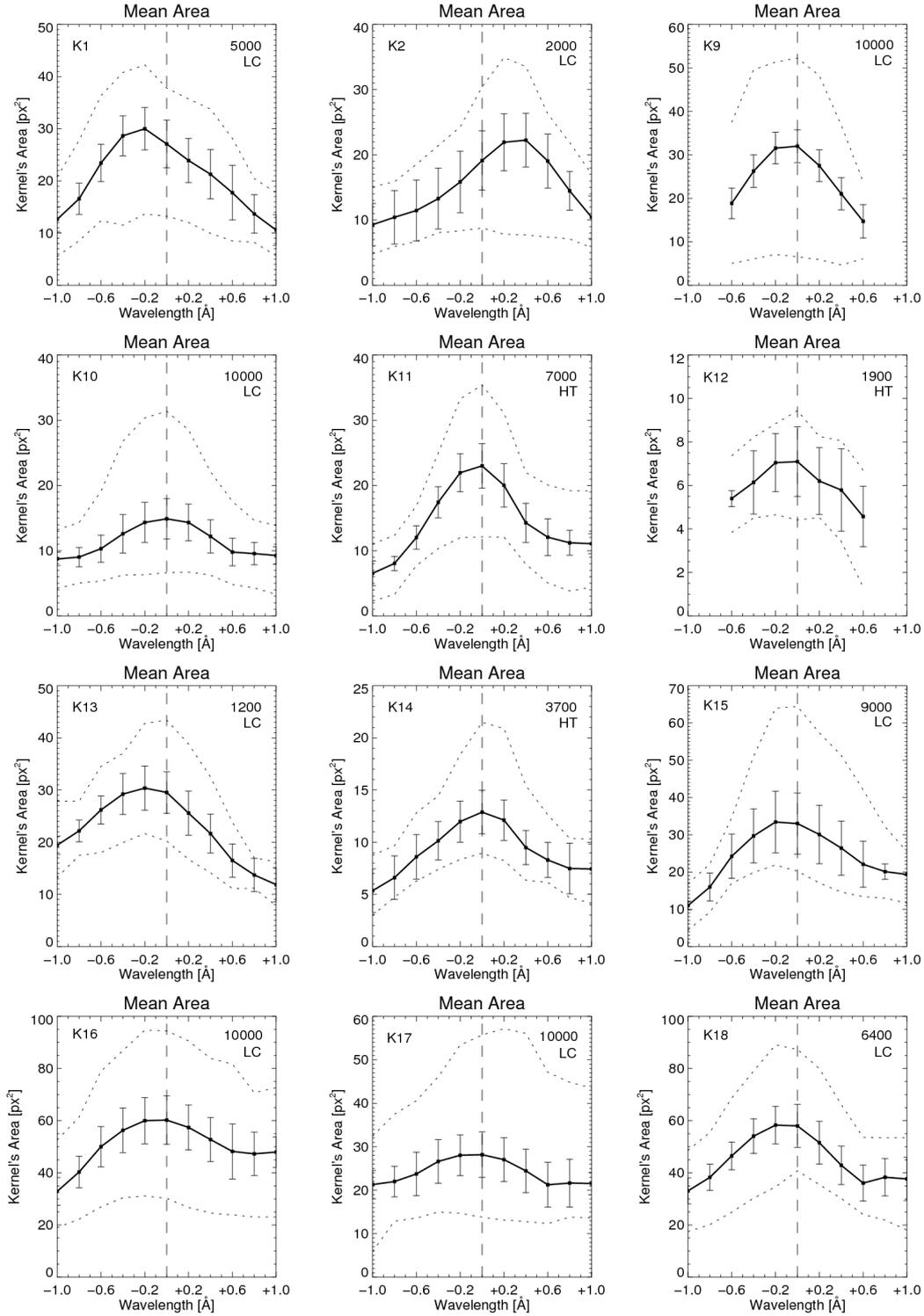

**Figure 3** Mean (solid line), maximum and minimum (dotted lines) instant areas of K1, K2 and K9-K18 flaring kernels, measured in 11 wavelengths, up to ±0.1 nm from the Hα line centre. The numbers of the analyzed (and averaged) measurements for each wavelength separately are given in the upper-right corners of the relevant figures. HT means data collected with the *Horizontal Telescope*, LC with the *Large Coronagraph*. The areas of the kernels are given in square pixels (in the centre of the solar disk 1 px$^2$ = 3.1 arcsec$^2$ (5×10$^6$ km$^2$) when observed with the HT and 1 px$^2$ = 1.1 arcsec$^2$ (~6×10$^5$ km$^2$) when observed with the LC, respectively). Error bars show standard deviations of the temporal variations.

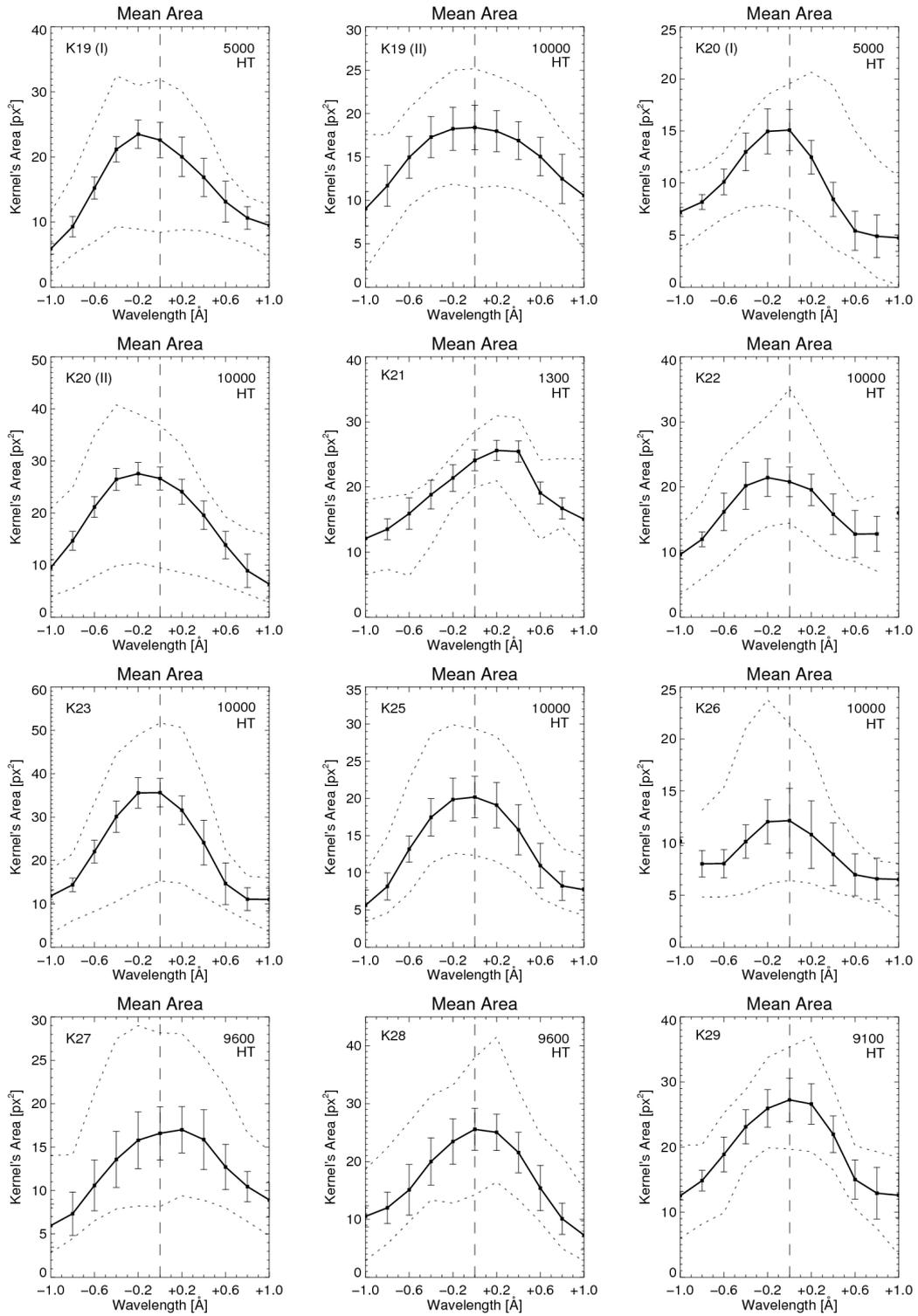

**Figure 4** Mean (solid line), maximum and minimum (dotted lines) instant areas of the K19-K23 and K25-K29 flaring kernels. Other details are the same as in Figure 3.

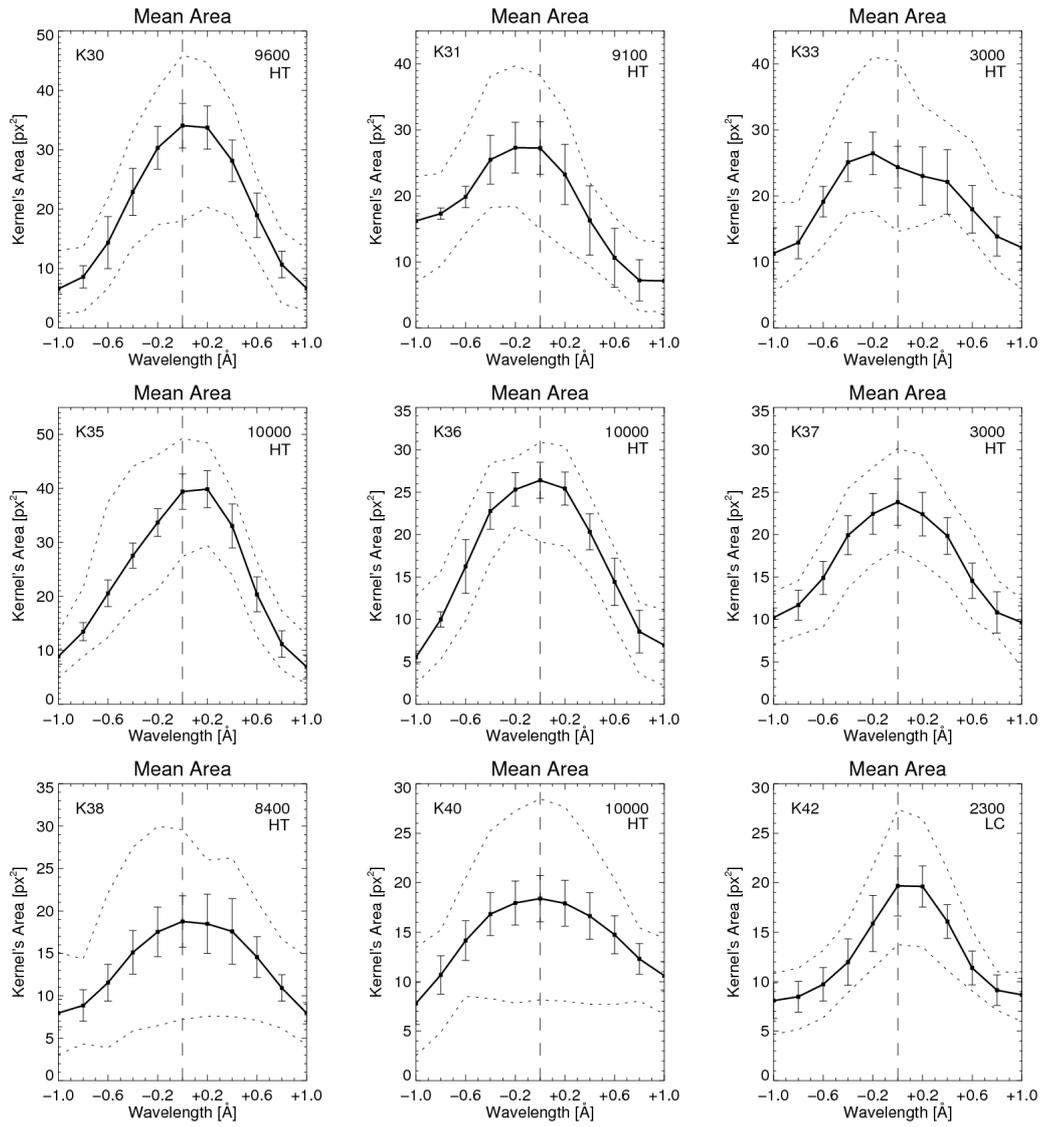

**Figure 5** Mean (solid line), maximum and minimum (dotted lines) instant areas of the K30, K31, K33, K35-K38, K40 and K42 flaring kernels. Other details are the same as in Figure 3.